\documentclass[doc]{apa6}
\usepackage{amsmath}
\usepackage{amsfonts}
\usepackage{amssymb}
\usepackage{listings}
\makeatletter
\def\Let@{\def\\{\notag\math@cr}}
\makeatother
\usepackage{graphicx}
\usepackage[utf8]{inputenc}
\usepackage[T1]{fontenc}

\usepackage[utf8]{inputenc}
\usepackage{amsmath,amssymb}
\newcommand{\textVerb}[1]{\texttt{\mbox{#1}}}
\usepackage{subcaption}
\usepackage{url}

\usepackage[american]{babel}
\usepackage{csquotes}
\usepackage{natbib}
\usepackage{caption}
\title{Estimating Psychological Networks and their Accuracy: A Tutorial Paper}

\shorttitle{Estimating Psychological Networks and their Accuracy}
\author{Sacha Epskamp, Denny Borsboom and Eiko I.\ Fried}

\affiliation{Department of Psychology, University of Amsterdam}
\abstract{
The usage of \emph{psychological networks} that conceptualize behaviour as a complex interplay of psychological and other components has gained increasing popularity in various research fields. While prior publications have tackled the topics of estimating and interpreting such networks, little work has been conducted to check how \emph{accurate} (i.e., prone to sampling variation) networks are estimated, and how \emph{stable} (i.e., interpretation remains similar with less observations) inferences from the network structure (such as centrality indices) are. In this tutorial paper, we aim to introduce the reader to this field and tackle the problem of accuracy under sampling variation. We first introduce the current state-of-the-art of network estimation. Second, we provide a rationale why researchers should investigate the accuracy of psychological networks. Third, we describe how bootstrap routines can be used to (A) assess the accuracy of estimated network connections, (B) investigate the stability of centrality indices, and (C) test whether network connections and centrality estimates for different variables differ from each other. We introduce two novel statistical methods: for (B) the \emph{correlation stability coefficient}, and for (C) the \emph{bootstrapped difference test} for edge-weights and centrality indices. We conducted and present simulation studies to assess the performance of both methods. Finally, we developed the free R-package \emph{bootnet} that allows for estimating psychological networks in a  generalized framework in addition to the proposed bootstrap methods. We showcase \emph{bootnet} in a tutorial, accompanied by R syntax, in which we analyze a dataset of 359 women with posttraumatic stress disorder available online.}
\begin{document}

\maketitle

\raggedbottom

\section{Introduction}

In the last five years, network research has gained substantial attention in psychological sciences \citep{borsboom2013network,cramer2010comorbidity}. In this field of research, psychological behavior is conceptualized as a complex interplay of psychological and other components. To portray a potential structure in which these components interact, researchers have made use of \emph{psychological networks}. Psychological networks consist of nodes representing observed variables, connected by edges representing statistical relationships. This methodology has gained substantial footing and has been used in various different fields of psychology, such as clinical psychology (e.g., \citealt{boschloo2015, fried2015, mcnally2015mental,forbush2016application}), psychiatry (e.g., \citealt{isvoranu,isvoranua, van2015association}), personality research (e.g., \citealt{costantini2015state,costantini2015development,cramer2012dimensions}), social psychology (e.g., \citealt{dalege2016toward}), and quality of life research (\citealt{kossakowski2015}). 

These analyses typically involve two steps: (1) estimate a statistical model on data, from which some parameters can be represented as a weighted network between observed variables, and (2), analyze the weighted network structure using measures taken from graph theory \citep{newman2010} to infer, for instance, the most central nodes.\footnote{An introduction on the interpretation and inference of network models has been included in the supplementary materials.} Step 1 makes psychological networks strikingly different from network structures typically used in graph theory, such as power grids \citep{watts1998}, social networks \citep{wasserman1994} or ecological networks \citep{barzel2009} in which nodes represent entities (e.g., airports, people, organisms) and connections are generally observed and known (e.g., electricity lines, friendships, mutualistic relationships). In psychological networks, the strength of connection between two nodes is a parameter \emph{estimated} from data. With increasing sample size, the parameters will be more accurately estimated (close to the true value). However, in the limited sample size psychological research typically has to offer, the parameters may not be estimated accurately, and in such cases, interpretation of the network and any measures derived from the network is questionable. Therefore, in estimating psychological networks, we suggest a third step is crucial: (3) assessing the accuracy of the network parameters and measures. 

\begin{figure*}
\centering
 \begin{subfigure}[b]{0.45\textwidth}
	  	\includegraphics[width=1\textwidth,page=1]{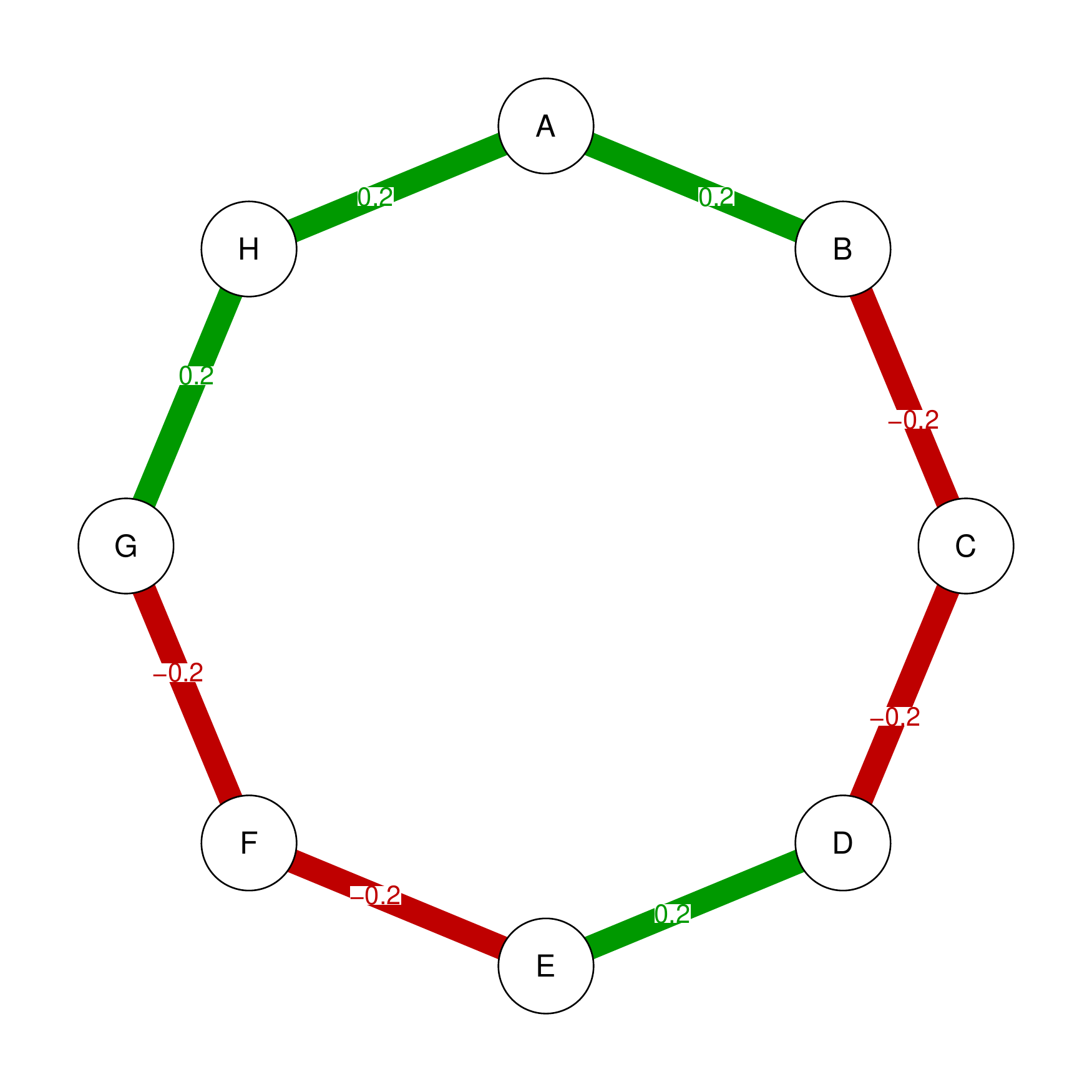}
        %\caption{Network structure\label{bootnet:fig:1a}}
    \end{subfigure}
 \begin{subfigure}[b]{0.45\textwidth}
	  	\includegraphics[width=1\textwidth,page=2]{Figure1.pdf}
        %\caption{Centrality indices\label{bootnet:fig:1b}}
    \end{subfigure}
\caption{Simulated network structure (left panel) and the importance of each node quantified in \emph{centrality indices} (right panel). The simulated network is a chain network in which each edge has the same absolute strength. The network model used was a Gaussian graphical model in which each edge represents partial correlation coefficients between two variables after conditioning on all other variables.}
\label{bootnet:fig:1}
\end{figure*}

\begin{figure*}
\centering
\begin{subfigure}[b]{0.45\textwidth}
	  	\includegraphics[width=1\textwidth,page=1]{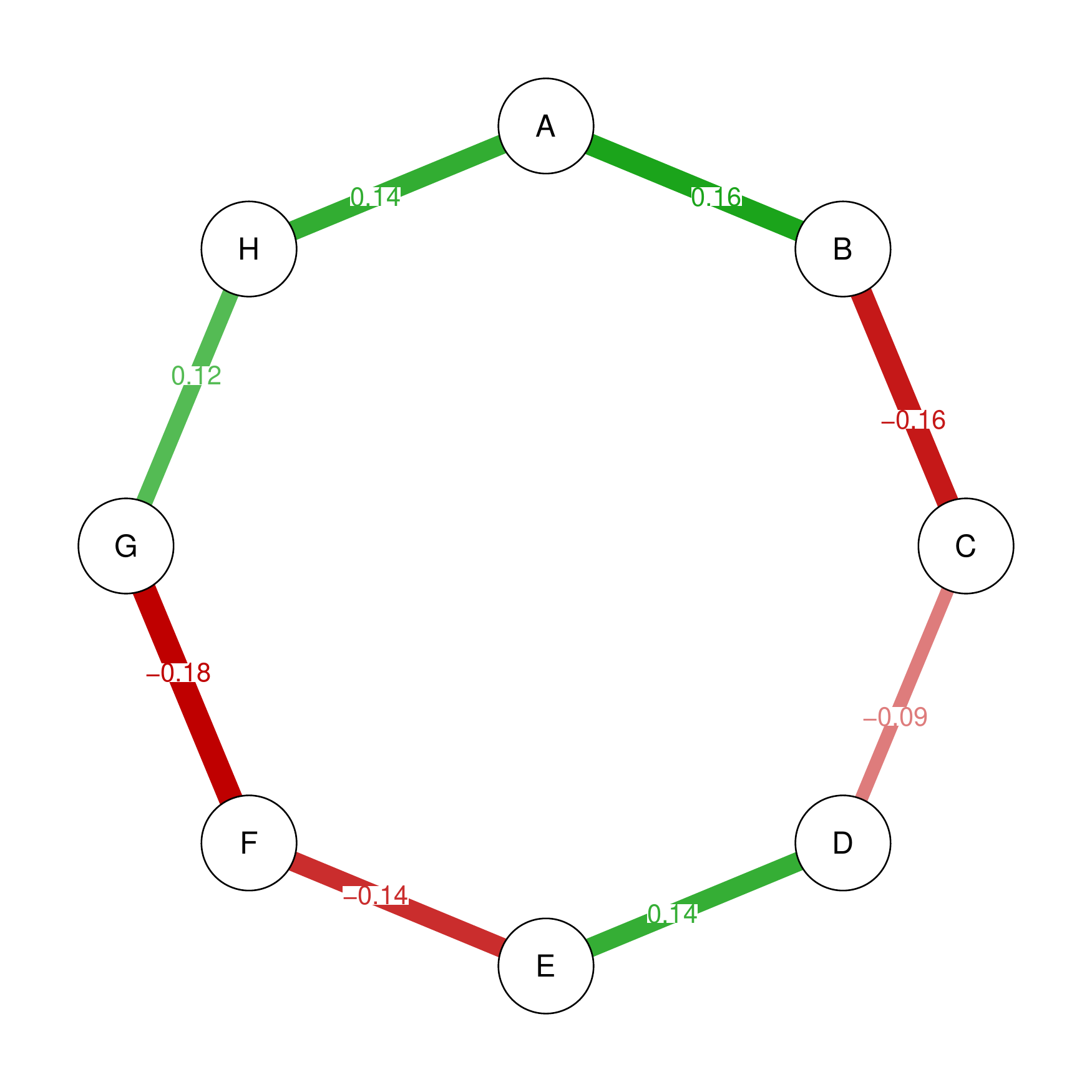}
       % \caption{Network structure\label{bootnet:fig:2a}}
    \end{subfigure}
 \begin{subfigure}[b]{0.45\textwidth}
	  	\includegraphics[width=1\textwidth,page=2]{Figure2.pdf}
       % \caption{Centrality indices\label{bootnet:fig:2b}}
    \end{subfigure}     
\caption{Estimated network structure based on a sample of 500 people simulated using the true model shown in Figure~\ref{bootnet:fig:1} (left panel) and computed centrality indices (right panel). Centrality indices are shown as standardized $z$-scores. Centrality indices show that nodes B and C are the most important nodes, even though the true model does not differentiate in importance between nodes.}
\label{bootnet:fig:2}
\end{figure*}

To highlight the importance of accuracy analysis in psychological networks, consider Figure~\ref{bootnet:fig:1} and Figure~\ref{bootnet:fig:2}. Figure~\ref{bootnet:fig:1} (left panel) shows a simulated network structure of 8 nodes in which each node is connected to two others in a \emph{chain network}. The network model used is a Gaussian graphical model \citep{lauritzen1996graphical}, in which nodes represent observed variables and edges represent \emph{partial correlation coefficients} between two variables after conditioning on all other variables in the dataset. A typical way of assessing the importance of nodes in this network is to \emph{centrality indices} of the network structure \citep{costantini2015state,newman2010,opsahl2010node}. Three such measures are \emph{node strength}, quantifying how well a node is directly connected to other nodes, \emph{closeness}, quantifying how well a node is indirectly connected to other nodes, and \emph{betweenness}, quantifying how important a node is in the average path between two other nodes. Figure~\ref{bootnet:fig:1} (right panel) shows the centrality indices of the true network:  all indices are exactly equal. We simulated a dataset of 500 individuals (typically regarded a moderately large sample size in psychology) using the network in Figure~\ref{bootnet:fig:1} and estimated a network structure based on the simulated data (as further described below). Results are presented in Figure~\ref{bootnet:fig:2}; this is the observed network structure that researchers are usually faced with, without knowing the true network structure. Of note, this network closely resembles the true network structure.\footnote{Penalized maximum likelihood estimation used in this analysis typically leads to slightly lower parameter estimates on average. As a result, the absolute edge-weights in Figure 2 are all closer to zero than the absolute edge-weights in the true network in Figure 1.} As can be seen in Figure~\ref{bootnet:fig:2} (right panel), however, centrality indices of the estimated network \emph{do} differ from each other. Without knowledge on how \emph{accurate} the centrality of these nodes are estimated, a researcher might in this case falsely conclude that node F (based on strength) and G and H (based on closeness and betweenness) play a much more important role in the network than other nodes.

Only few analyses so far have taken accuracy into account (e.g., \citealt{fried2016}), mainly because the meth\-od\-ology has not yet been worked out. This problem of accuracy is omnipresent in statistics. Imagine researchers employ a regression analysis to examine three predictors of depression severity, and identify one strong, one weak, and one unrelated regressor. If removing one of these three regressors, or adding a fourth one, substantially changes the regression coefficients of the other regressors, results are unstable and depend on specific decisions the researchers make, implying a problem of accuracy. The same holds for psychological networks. Imagine in a network of psychopathological symptoms that we find that symptom A has a much higher node strength than symptom B in a psychopathological network, leading to the clinical interpretation that A may be a more relevant target for treatment than the peripheral symptom B \citep{fried2016}. Clearly, this interpretation relies on the assumption that the centrality estimates are indeed different from each other. Due to the current uncertainty, there is the danger to obtain network structures sensitive to specific variables included, or sensitive to specific estimation methods. This poses a major challenge, especially when substantive interpretations such as treatment recommendations in the psychopathological literature, or the generalizability of the findings, are important. The current replication crisis in psychology \citep{opensciencecollaboration2015} stresses the crucial importance of obtaining robust results, and we want the emerging field of psychopathological networks to start off on the right foot.

The remainder of the article is structured into three sections. In the first section, we give a brief overview of often used methods in estimating psychological networks, including an overview of open-source software packages that implement these methods available in the statistical programming environment R \citep{R}. In the second section, we outline a methodology to assess the accuracy of psychological network structures that includes three steps: (A) estimate confidence intervals (CIs) on the edge-weights, (B) assess the \emph{stability} of centrality indices under observing subsets of cases, and (C) test for significant differences between edge-weights and centrality indices. We introduce the freely available R package, \emph{bootnet}\footnote{CRAN link: \url{http://cran.r-project.org/package=bootnet}\\Github link (developmental): \url{http://www.github.com/SachaEpskamp/bootnet}}, that can be used both as a generalized framework to estimate various different network models as well as to conduct the accuracy tests we propose. We demonstrate the package's functionality of both estimating networks and checking their accuracy in a step-by-step tutorial using a dataset of 359 women with post-traumatic stress disorder (PTSD; \citealt{hien2009multisite}) that can be downloaded from the Data Share Website of the National Institute on Drug Abuse. Finally, in the last section, we show the performance of the proposed methods for investigating accuracy in three simulations studies. It is important to note that the focus of our tutorial is on cross-sectional network models that can readily be applied to many current psychological datasets. Many sources have already outlined the interpretation of probabilistic network models (e.g., \citealt{epskampPsychometrika,koller2009probabilistic,lauritzen1996graphical}), as well as network inference techniques, such as centrality measures, that can be used once a network is obtained (e.g., \citealt{costantini2015state,kolaczyk2009statistical,newman2004analysis,sporns2004organization}). 

To make this tutorial stand-alone readable for psychological researchers, we included a detailed description of how to interpret psychological network models as well as an overview of network measures in the supplementary materials. We hope that this tutorial will enable researchers to gauge the accuracy and certainty of the results obtained from network models, and to provide editors, reviewers, and readers of psychological network papers the possibility to better judge whether substantive conclusions drawn from such analyses are defensible.

% CHANGE TO: To make this dissertation stand-alone readable, a description of how to interpret network models as well as an overview of network measures has been included in Chapter~\ref{introduction}\footnote{This text is adapted from the same paper on which the current chapter is based, which I included in the introduction rather than the current chapter as it provides key background knowledge to many chapters of this dissertation.}.

\section{Estimating Psychological Networks}

As described in more detail in the supplementary materials, a popular network model to use in estimating psychological networks is a pairwise Markov Random Field (PMRF; \citealt{costantini2015state,van2014new}), on which the present paper is focused. It should be noted, however, that the described methodology could be applied to other network models as well. A PMRF is a network in which nodes represent variables, connected by undirected edges (edges with no arrowhead) indicating conditional dependence between two variables; two variables that are not connected are independent after conditioning on other variables. When data are multivariate normal, such a conditional independence would correspond to a partial correlation being equal to zero. Conditional independencies are also to be expected in many causal structures \citep{pearl2000causality}. In cross-sectional observational data, causal networks (e.g. directed networks) are hard to estimate without stringent assumptions (e.g., no feedback loops). In addition, directed networks suffer from a problem of many equivalent models (e.g., a network $A \rightarrow B$ is not statistically distuinghuisable from a network $A \leftarrow B$; \citealt{maccallum1993problem}). PMRFs, however, are well defined and have no equivalent models (i.e., for a given PMRF, there exists no other PMRF that describes exactly the same statistical independence relationships for the set of variables under consideration). Therefore, they facilitate a clear and unambiguous interpretation of the edge-weight parameters as strength of unique associations between variables, which in turn may highlight potential causal relationships. 

When the data are binary, the appropriate PRMF model to use is called the Ising model \citep{van2014new}, and requires binary data to be estimated. When the data follow a multivariate normal density, the appropriate PRMF model is called the Gaussian graphical model (GGM; \citealt{costantini2015state,lauritzen1996graphical}), in which edges can directly be interpreted as \emph{partial correlation coefficients}. The GGM requires an estimate of the covariance matrix as input,\footnote{While the GGM requires a covariance matrix as input, it is important to note that the model itself is based on the (possibly sparse) \emph{inverse} of the covariance matrix. Therefore, the network shown does not show marginal correlations (regular correlation coefficients between two variables). The inverse covariance matrix instead encodes \emph{partial} correlations.} for which polychoric correlations can also be used in case the data are ordinal \citep{qgraphsims}. For continuous data that are not normally distributed, a transformation can be applied (e.g., by using the \emph{nonparanormal transformation}; \citealt{liu2012high}) before estimating the GGM. Finally, mixed graphical models can be used to estimate a PMRF containing both continuous and categorical variables \citep{jonas2}.

\paragraph{Dealing with the problem of small N in psychological data} Estimating a PMRF features a severe limitation: the number of parameters to estimate grows quickly with the size of the network. In a $10$-node network, $55$ parameters ($10$ threshold parameters and $10 \times 9 / 2 = 45$ pairwise association parameters) need be estimated already. This number grows to $210$ in a network with $20$ nodes, and to $1275$ in a $50$-node network. To reliably estimate that many parameters, the number of observations needed typically exceeds the number available in characteristic psychological data. To deal with the problem of relatively small datasets, recent researchers using psychological networks have applied the `least absolute shrinkage and selection operator' (LASSO; \citealt{tibshirani1996regression}). This technique is a form of \emph{regularization}. The LASSO employs such a regularizing penalty by limiting the total sum of absolute parameter values---thus treating positive and negative edge-weights equally---leading many edge estimates to shrink to exactly zero and dropping out of the model. As such, the LASSO returns a \emph{sparse} (or, in substantive terms, conservative) network model: only a relatively small number of edges are used to explain the covariation structure in the data. Because of this sparsity, the estimated models become more interpretable. The LASSO utilizes a tuning parameter to control the degree to which regularization is applied. This tuning parameter can be selected by minimizing the Extended Bayesian Information Criterion (EBIC; \citealt{chen2008EBIC}). Model selection using the EBIC has been shown to work well in both estimating the Ising model \citep{foygel2014high,van2014new} and the GGM \citep{foygel2010extended}. The remainder of this paper focuses on the GGM estimation method proposed by Foygel \& Drton (\citeyear{foygel2010extended}; see also \citealt{primerpaper}, for a detailed introduction of this method for psychological researchers).

Estimating regularized networks in R is straightforward. For the Ising model, LASSO estimation using EBIC has been implemented in the \emph{IsingFit} package \citep{van2014new}. For GGM networks, a well-established and fast algorithm for estimating LASSO regularization is the \emph{graphical LASSO} (glasso; \citealt{friedman2008sparse}), which is implemented in the package \emph{glasso} \citep{glasso}. The \emph{qgraph} package utilizes \emph{glasso} in combination with EBIC model selection to estimate a regularized GGM. Alternatively, the \emph{huge} \citep{huge} and \emph{parcor} \citep{parcor} packages implement several regularization methods---including also glasso with EBIC model selection---to estimate a GGM. Finally, mixed graphical models have been implemented in the \emph{mgm} package \citep{mgm}.

\section{Network Accuracy}

The above description is an overview of the current state of network estimation in psychology. While network inference is typically performed by assessing edge strengths and node centrality, little work has been done in investigating how accurate these inferences are. This section will outline methods that can be used to gain insights into the accuracy of edge weights and the stability of centrality indices in the estimated network structure. We outline several methods that should routinely be applied after a network has been estimated. These methods will follow three steps: (A) estimation of the accuracy of edge-weights, by drawing bootstrapped CIs; (B) investigating the stability of (the order of) centrality indices after observing only portions of the data; and (C) performing bootstrapped difference tests between edge-weights and centrality indices to test whether these differ significantly from each other. We introduced these methods in decreasing order of importance: while (A) should always be performed, a researcher not interested in centrality indices might not perform other steps, whereas a researcher not interested in testing for differences might only perform (A) and (B). 
 studies have been conducted to assess the performance of these methods, which are reported in a later section in the paper.

\subsection{Edge-weight Accuracy}

To assess the variability of edge-weights, we can estimate a CI: in $95\%$ of the cases such a CI will contain the true value of the parameter. To construct a CI, we need to know the \emph{sampling distribution} of the statistic of interest. While such sampling distributions can be difficult to obtain for complicated statistics such as centrality measures, there is a straight-forward way of constructing CIs many statistics: \emph{bootstrapping} \citep{efron1979bootstrap}. Bootstrapping involves repeatedly estimating a model under sampled or simulated data and estimating the statistic of interest.  Following the bootstrap, a $1-\alpha$ CI can be approximated by taking the interval between quantiles $1/2 \alpha$ and $1 - 1/2 \alpha$ of the bootstrapped values. We term such an interval a \emph{bootstrapped CI}. Bootstrapping edge-weights can be done in two ways: using non-parametric bootstrap and parametric bootstrap \citep{bollen1992bootstrapping}. In \emph{non-parametric} bootstrapping, observations in the data are resampled with replacement to create new plausible datasets, whereas \emph{parametric} bootstrapping samples new observations from the parametric model that has been estimated from the original data; this creates a series of values that can be used to estimate the sampling distribution. Bootstrapping can be applied as well to LASSO regularized statistics \citep{hastie2015statistical}. 

With $N_B$ bootstrap samples, at maximum a CI with $\alpha = 2/N_B$ can be formed. In this case, the CI equals the range of bootstrapped samples and is based on the two most extreme samples (minimum and maximum). As such, for a certain level of $\alpha$ at the very least $2/\alpha$ bootstrap samples are needed. It is recommended however to use more bootstrap samples to improve consistency of results. The estimation of quantiles is not trivial and can be done using various methods \citep{hyndman1996sample}.  In unreported simulation studies available on request, we found that the default quantile estimation method used in R (type 7; \citealt{quantile7}) constructed CIs that were too small when samples are normally or uniformly distributed, inflating $\alpha$. We have thus changed the method to type 6, described in detail by \citet{hyndman1996sample}, which resulted in CIs of proper width in uniformly distributed samples, and slightly wider CIs when samples were distributed normally. Simulation studies below that use type 6 show that this method allows for testing of significant differences at the correct $\alpha$ level.

Non-parametric bootstrapping can always be applied, whereas parametric bootstrapping requires a parametric model of the data. When we estimate a GGM, data can be sampled by sampling from the multivariate normal distribution through the use of the R package \emph{mvtnorm} \citep{genz2008mvtnorm}; to sample from the Ising model, we have developed the R package \emph{IsingSampler} \citep{IsingSampler}. Using the GGM model, the parametric bootstrap samples continuous multivariate normal data---an important distinction from ordinal data if the GGM was estimated using polychoric correlations. Therefore, we advise the researcher to use the non-parametric bootstrap when handling ordinal data. Furthermore, when LASSO regularization is used to estimate a network, the edge-weights are on average made smaller due to shrinkage, which biases the parametric bootstrap. The non-parametric bootstrap is in addition fully data-driven and requires no theory, whereas the parametric bootstrap is more theory driven. As such, we will only discuss the non-parametric bootstrap in this paper and advice the researcher to only use parametric bootstrap when no regularization is used and if the non-parametric results prove unstable or to check for correspondence of bootstrapped CIs between both methods.

It is important to stress that the bootstrapped results should \emph{not} be used to test for significance of an edge being different from zero. While unreported simulation studies showed that observing if zero is in the bootstrapped CI does function as a valid null-hypothesis test (the null-hypothesis is rejected less than $\alpha$ when it is true), the utility of testing for significance in LASSO regularized edges is questionable. In the case of partial correlation coefficients, without using LASSO the sampling distribution is well known and $p$-values are readily available. LASSO regularization aims to estimate edges that are not needed to be exactly zero. Therefore, observing that an edge is not set to zero already indicates that the edge is sufficiently strong to be included in the model. In addition, as later described in this paper, applying a correction for multiple testing is not feasible, In sum, the edge-weight bootstrapped CIs should not be interpreted as significance tests to zero, but only to show the accuracy of edge-weight estimates and to compare edges to one-another. 

When the bootstrapped CIs are wide, it becomes hard to interpret the strength of an edge. Interpreting the presence of an edge, however, is not affected by large CIs as the LASSO already performed model selection. In addition, the sign of an edge (positive or negative) can also be interpreted regardless of the width of a CI as the LASSO rarely retains an edge in the model that can either be positive or negative. As centrality indices are a direct function of edge weights, large edge weight CIs will likely result in a poor accuracy for centrality indices as well. However, differences in centrality indices can be accurate even when there are large edge weight CIs, and vice-versa; and there are situations where differences in centrality indices can also be hard to interpret even when the edge weight CIs are small (for example, when centrality of nodes do not differ from one-another). The next section will detail steps to investigate centrality indices in more detail.

\subsection{Centrality Stability}

While the bootstrapped CIs of edge-weights can be constructed using the bootstrap, we discovered in the process of this research that constructing CIs for centrality indices is far from trivial. As discussed in more detail in the supplementary materials, both estimating centrality indices based on a sample and bootstrapping centrality indices result in biased sampling distributions, and thus the bootstrap cannot readily be used to construct true $95\%$ CIs even without regularization. To allow the researcher insight in the accuracy of the found centralities, we suggest to investigate the stability of the order of centrality indices based on \emph{subsets} of the data. With \emph{stability}, we indicate if the order of centrality indices remains the same after re-estimating the network with less cases or nodes. A \emph{case} indicates a single observation of all variables (e.g., a person in the dataset) and is represented by \emph{rows} of the dataset. Nodes, on the other hand, indicate \emph{columns} of the dataset. Taking subsets of cases in the dataset employs the so called \emph{$m$ out of $n$ bootstrap}, which is commonly used to remediate problems with the regular bootstrap \citep{chernick2011bootstrap}. Applying this bootstrap for various proportions of cases to drop can be used to assess the correlation between the original centrality indices and those obtained from subsets. If this correlation completely changes after dropping, say, $10\%$ of the cases, then interpretations of centralities are prone to error. We term this framework the \emph{case-dropping subset bootstrap}. Similarly, one can opt to investigate the stability of centrality indices after dropping nodes from the network (\emph{node-dropping subset bootstrap}; \citealt{costenbader2003stability}), which has also been implemented in \emph{bootnet} but is harder to interpret (dropping $50\%$ of the nodes leads to entirely different network structures). As such, we only investigate stability under case-dropping, while noting that the below described methods can also be applied to node-dropping.

To quantify the stability of centrality indices using subset bootstraps, we propose a measure we term the \emph{correlation stability coefficient}, or short, the $CS$-coefficient. Let $CS( \mathrm{cor} = 0.7 )$ represent the maximum proportion of cases that can be dropped, such that with $95\%$ probability the correlation between original centrality indices and centrality of networks based on subsets is $0.7$ or higher. The value of $0.7$ can be changed according to the stability a researcher is interested in, but is set to $0.7$ by default as this value has classically been interpreted as indicating a very large effect in the behavioral sciences \citep{cohen1977statistical}. The simulation study below showed that to interpret centrality differences the $CS$-coefficient should not be below $0.25$, and preferably above $0.5$. While these cutoff scores emerge as recommendations from this simulation study, however, they are somewhat arbitrary and should not be taken as definite guidelines.

\subsection{Testing for Significant Differences}

In addition to investigating the accuracy of edge weights and the stability of the order of centrality, researchers may wish to know whether a specific edge $A$--$B$ is significantly larger than another edge $A$--$C$, or whether the centrality of node A is significantly larger than that of node B. To that end, the bootstrapped values can be used to test if two edge-weights or centralities significantly differ from one-another. This can be done by taking the \emph{difference} between bootstrap values of one edge-weight or centrality and another edge-weight or centrality, and constructing a bootstrapped CI around those difference scores. This allows for a null-hypothesis test if the edge-weights or centralities differ from one-another by checking if zero is in the bootstrapped CI \citep{chernick2011bootstrap}. We term this test the \emph{bootstrapped difference test}.

As the bootstraps are functions of complicated estimation methods, in this case LASSO regularization of partial correlation networks based on polychoric correlation matrices, we assessed the performance of the bootstrapped difference test for both edge-weights and centrality indices in two simulation studies below. The edge-weight bootstrapped difference test performs well with Type~I error rate close to the significance level ($\alpha$), although the test is slightly conservative at low sample sizes (i.e, due to edge-weights often being set to zero, the test has a Type~I error rate somewhat less than $\alpha$). When comparing two centrality indices, the test also performs as a valid, albeit somewhat conservative, null-hypothesis test with Type~I error rate close to or less than $\alpha$. However, this test does feature a somewhat lower level of power in rejecting the null-hypothesis when two centralities do differ from one-another.

A null-hypothesis test, such as the bootstrapped difference test, can only be used as evidence that two values differ from one-another (and even then care should be taken in interpreting its results; e.g., \citealt{Cohen1994}). \emph{Not} rejecting the null-hypothesis, however, does not necessarily constitute evidence for the null-hypothesis being true \citep{wagenmakers2007practical}. The slightly lower power of the bootstrapped difference test implies that, at typical sample sizes used in psychological research, the test will tend to find fewer significant differences than actually exist at the population level. Researchers should therefore not routinely take nonsignificant centralities as evidence for centralities being equal to each other, or for the centralities not being accurately estimated. Furthermore, as described below, applying a correction for multiple testing is not feasible in practice. As such, we advise care when interpreting the results of bootstrapped difference tests.

\paragraph{A note on multiple testing} The problem of performing multiple significance tests is well known in statistics. When one preforms two tests, both at $\alpha = 0.05$, the probability of finding at least \emph{one} false significant result (Type~I error) is \emph{higher} than $5\%$. As a result, when performing a large number of significance tests, even when the null-hypothesis is true in all tests one would likely find several significant results purely by chance. To this end, researchers often apply a \emph{correction for multiple testing}. A common correction is the `Bonferroni correction' \citep{bland1995multiple}, in which $\alpha$ is divided by the number of tests. To test, for example, differences between all edge-weights of a 20-node network requires $17{,}955$ tests, leading to a Bonferroni corrected significance level of $0.000003$.\footnote{One might instead only test for difference in edges that were estimated to be non-zero with the LASSO. However, doing so still often leads to a large number of tests.} Testing at such a low significance level is \emph{not} feasible with the proposed bootstrap methods, for three reasons:
\begin{enumerate}
\item The distribution of such LASSO regularized parameters is far from normal \citep{potscher2009distribution}, and as a result approximate $p$-values cannot be obtained from the bootstraps. This is particularly important for extreme significance levels that might be used when one wants to test using a correction for multiple testing. It is for this reason that this paper does not mention bootstrapping $p$-values and only investigates null-hypothesis tests by using bootstrapped CIs.
\item When using bootstrapped CIs with $N_B$ bootstrap samples, the widest interval that can be constructed is the interval between the two most extreme bootstrap values, corresponding to $\alpha = 2/N_B$. With $1{,}000$ bootstrap samples, this corresponds to $\alpha = 0.002$. Clearly, this value is much higher than $0.000003$ mentioned above. Taking the needed number of bootstrap samples for such small significance levels is computationally challenging and not feasible in practice. 
\item In significance testing there is always interplay of Type~I and Type~II error rates: when one goes down, the other goes up. As such, reducing the Type I error rate increases the Type~II error rate (not rejecting the null when the alternative hypothesis is true), and thus reduces statistical power. In the case of $\alpha = 0.000003$, even if we could test at this significance level, we would likely find no significant differences due to the low statistical power. 
\end{enumerate}
As such, Bonferroni corrected difference tests are still a topic of future research.

\subsection{Summary}

In sum, the non-parametric (resampling rows from the data with replacement) bootstrap can be used to assess the \emph{accuracy} of network estimation, by investigating the sampling variability in edge-weights, as well as to test if edge-weights and centrality indices significantly differ from one-another using the bootstrapped difference test. Case-dropping subset bootstrap (dropping rows from the data), on the other hand, can be used to assess the \emph{stability} of centrality indices, how well the order of centralities are retained after observing only a subset of the data. This stability can be quantified using the $CS$-coefficient. The R code in the supplementary materials show examples of these methods on the simulated data in Figure~\ref{bootnet:fig:1} and Figure~\ref{bootnet:fig:2}. As expected from Figure~\ref{bootnet:fig:1}, showing that the true centralities did not differ, bootstrapping reveals that none of the centrality indices in Figure 2 significantly differ from one-another. In addition, node strength ($CS(\mathrm{cor}=0.7) = 0.08$), closeness ($CS(\mathrm{cor}=0.7) = 0.05$) and betweenness ($CS(\mathrm{cor}=0.7) = 0.05$) were far below the thresholds that we would consider stable. Thus, the novel bootstrapping methods proposed and implemented here showed that the differences in centrality indices presented in Figure~\ref{bootnet:fig:2} were not interpretable as true differences.

\section{Tutorial}

In this section, we showcase the functionality of the \emph{bootnet} package for estimating network structures and assessing their accuracy. We do so by analyzing a dataset ($N = 359$) of women suffering from posttraumatic stress disorder (PTSD) or sub-threshold PTSD. The \emph{bootnet} package includes the bootstrapping methods, $CS$-coefficient and bootstrapped difference tests as described above. In addition, \emph{bootnet} offers a wide range of plotting methods. After estimating nonparametric bootstraps, \emph{bootnet} produces plots that show the bootstrapped CIs of edge-weights or which edges and centrality indices significantly differ from one-another. After estimating subset bootstrap, \emph{bootnet} produces plots that show the correlation of centrality indices under different levels of subsetting \citep{costenbader2003stability}. In addition to the correlation plot, \emph{bootnet} can be used to plot the average estimated centrality index for each node under different sampling levels, giving more detail on the order of centrality under different subsetting levels.

With \emph{bootnet}, users can not only perform accuracy and stability tests, but also flexibly estimate a wide variety of network models in R. The estimation technique can be specified as a \emph{chain} of R commands, taking the data as input and returning a network as output. In bootnet, this chain is broken in several phases: data preparation (e.g., correlating or binarizing), model estimation (e.g., glasso) and network selection. The \emph{bootnet} package has several \emph{default sets}, which can be assigned using the \textVerb{default} argument in several functions. These default sets can be used to easily specify the most commonly used network estimation procedures. Table~\ref{bootnet:table:2} gives an overview of the default sets and the corresponding R functions called.\footnote{The notation makes use of notation introduced by the \emph{magrittr} R package \citep{magrittr}}

\begin{table*}
\centering
{\small
\begin{tabular}{ll}
Default set &	R chain \\
\hline
\textVerb{EBICglasso}	 &  \verb|Data %>% qgraph::cor_auto %>% qgraph::EBICglasso| \\
\textVerb{pcor} & \verb|Data %>% qgraph::cor_auto %>% corpcor::cor2pcor| \\
\textVerb{IsingFit}  &	\verb|Data %>% bootnet::binarize %>% IsingFit::IsingFit| \\
\textVerb{IsingLL}  &	\verb|Data %>% bootnet::binarize %>%|  \\
 & \verb|IsingSampler::EstimateIsing(method = “ll”)| \\
 \textVerb{huge} & \verb|Data %>% as.matrix %>% na.omit %>% huge::huge.npn %>%| \\
 & \verb|huge::huge(method = “glasso”) %>%| \\
 & \verb|huge::huge.select(criterion = “ebic”)| \\
\textVerb{adalasso} & \verb|Data %>% parcor::adalasso.net| \\
\hline
\end{tabular}
}
\caption{R chains to estimate network models from data. \normalfont{The default sets \texttt{"EBICglasso"}, \texttt{"pcor"}, \texttt{"huge"} and \texttt{"adalasso"} estimate a Gaussian graphical model and the default sets \texttt{"IsingFit"} and \texttt{"IsingLL"} estimate the Ising model. The notation \texttt{package::function} indicates that the function after the colons comes from the package before the colons. Chains are schematically represented using \emph{magrittr} chains: Whatever is on the left of \texttt{\%>\%} is used as first argument to the function on the right of this operator. Thus, the first chain corresponding to \texttt{"EBICglasso"} can also be read as \texttt{qgraph::EBICglasso(qgraph::cor\symbol{95}auto(Data))}.}}
\label{bootnet:table:2}
\end{table*}

\subsection{Example: Post-traumatic Stress Disorder}

To exemplify the usage of \emph{bootnet} in both estimating and investigating network structures, we use a dataset of 359 women enrolled in community-based substance abuse treatment programs across the United States (study title: Women's Treatment for Trauma and Substance Use Disorders; study number: NIDA-CTN-0015).\footnote{\url{https://datashare.nida.nih.gov/protocol/nida-ctn-0015}} All participants met the criteria for either PTSD or sub-threshold PTSD, according to the DSM-IV-TR \citep{APA2000}. Details of the sample, such as inclusion and exclusion criteria as well as demographic variables, can be found elsewhere \citep{hien2009multisite}. We estimate the network using the 17 PTSD symptoms from the PTSD Symptom Scale-Self
Report (PSS-SR; \citealt{foa1993reliability}). Participants rated the frequency of endorsing these symptoms on a scale ranging from 0 (not at all) to 3 (at least 4 or 5 times a week).

\begin{figure*}
\centering
\begin{subfigure}[b]{0.45\textwidth}
	  	\includegraphics[width=1\textwidth,page=1]{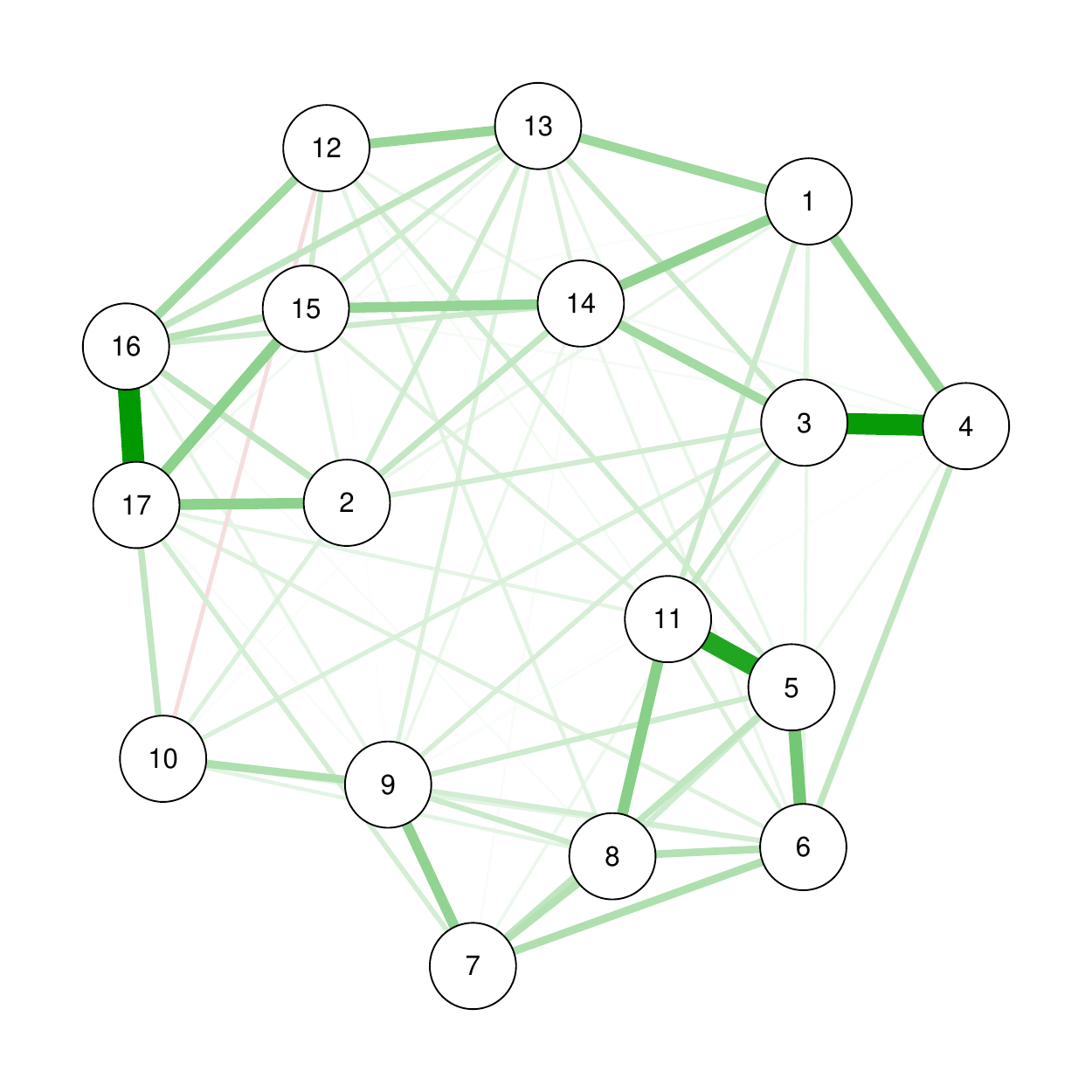}
 %       \caption{Network structure\label{bootnet:fig:neta}}
    \end{subfigure}
 \begin{subfigure}[b]{0.45\textwidth}
	  	\includegraphics[width=1\textwidth,page=2]{Figure3.pdf}
        %\caption{Centrality indices\label{bootnet:fig:netb}}
    \end{subfigure}  
\caption{Estimated network structure of 17 PTSD symptoms (left panel) and the corresponding centrality indices (right panel).  Centrality indices are shown as standardized $z$-scores. The network structure is a Gaussian graphical model, which is a network of partial correlation coefficients.}
\label{bootnet:fig:net}
\end{figure*}

\paragraph{Network estimation} Following the steps in Appendix~A, the data can be loaded into R in a data frame called \textVerb{Data}, which contains the frequency ratings at the baseline measurement point. We will estimate a Gaussian graphical model, using the graphical LASSO in combination with EBIC model selection as described above \citep{foygel2010extended}. This procedure requires an estimate of the variance-covariance matrix and returns a parsimonious network of \emph{partial correlation coefficients}. Since the PTSD symptoms are ordinal, we need to compute a polychoric correlation matrix as input. We can do so using the \textVerb{cor\_auto} function from the \emph{qgraph} package, which automatically detects ordinal variables and utilizes the R-package \textVerb{lavaan} \citep{lavaan} to compute polychoric (or, if needed, polyserial and Pearson) correlations. Next, the \textVerb{EBICglasso} function from the \emph{qgraph} package can be used to estimate the network structure, which uses the \emph{glasso} package for the actual computation \citep{glasso}. In \emph{bootnet}, as can be seen in Table~\ref{bootnet:table:2}, the \textVerb{"EBICglasso"}  default set automates this procedure. To estimate the network structure, one can use the \textVerb{estimateNetwork} function:
\begin{verbatim}
library("bootnet")	
Network <- estimateNetwork(Data, default = "EBICglasso")
\end{verbatim}
Next, we can plot the network using the plot method:
\begin{verbatim}
plot(Network, layout = "spring", labels = TRUE)	
\end{verbatim}
The \textVerb{plot} method uses \emph{qgraph} to plot the network. Figure~\ref{bootnet:fig:net} (left panel) shows the resulting network structure, which is parsimonious due to the LASSO estimation; the network only has 78 non-zero edges out of 136 possible edges. A description of the node labels can be seen in Table~\ref{bootnet:table:3}. Especially strong connections emerge among Node~3 (being jumpy) and Node~4 (being alert), Node~5 (cut off from people) and Node~11 (interest loss), and Node~16 (upset when reminded of the trauma) and Node~17 (upsetting thoughts/images). Other connections are absent, for instance between Node~7 (irritability) and Node~15 (reliving the trauma); this implies that these symptoms can be statistically independent when conditioning on all other symptoms (their partial correlation is zero) or that there was not sufficient power to detect an edge between these symptoms.

\begin{table}
\centering
\begin{tabular}{ll}
ID & Variable \\
\hline
1 & Avoid reminds of the trauma \\
2 & Bad dreams about the trauma \\
3 & Being jumpy or easily startled \\
4 & Being over alert \\
5 & Distant or cut off from people \\
6 & Feeling emotionally numb \\
7 & Feeling irritable \\
8 & Feeling plans won’t come true \\
9 & Having trouble concentrating \\
10 & Having trouble sleeping \\
11 & Less interest in activities \\
12 & Not able to remember \\
13 & Not thinking about trauma \\
14 & Physical reactions \\
15 & Reliving the trauma \\
16 & Upset when reminded of trauma \\
17 & Upsetting thoughts or images \\
\hline
\end{tabular}
\caption{Node IDs and corresponding symptom names of the 17 PTSD symptoms.}
\label{bootnet:table:3}
\end{table}

\paragraph{Computing centrality indices} To investigate centrality indices in the network, we can use the \textVerb{centralityPlot} function from the \emph{qgraph} package:
\begin{verbatim}
library("qgraph")
centralityPlot(Network)
\end{verbatim}
The resulting plot is shown in Figure~\ref{bootnet:fig:net} (right panel). It can be seen that nodes differ quite substantially in their centrality estimates. In the network, Node~17 (upsetting thoughts/images) has the highest strength and betweenness and Node~3 (being jumpy) has the highest closeness. However, without knowing the accuracy of the network structure and the stability of the centrality estimates, we cannot conclude whether the differences of centrality estimates are interpretable or not.

\paragraph{Edge-weight accuracy} The \textVerb{bootnet} function can be used to perform the bootstrapping methods described above. The function can be used in the same way as the \textVerb{estimateNetwork} function, or can take the output of the \textVerb{estimateNetwork} function to run the bootstrap using the same arguments. By default, the nonparametric bootstrap with $1{,}000$ samples will be used. This can be overwritten using the \textVerb{nBoots} argument, which is used below to obtain more smooth plots.\footnote{Using many bootstrap samples, such as the $2{,}500$ used here, might result in memory problems or long computation time. It is advisable to first use a small number of samples (e.g., 10) and then try more. The simulations below show that $1{,}000$ samples may often be sufficient.} The \textVerb{nCores} argument can be used to speed up bootstrapping and use multiple computer cores (here, eight cores are used):
\begin{verbatim}
boot1 <- bootnet(Network, nBoots = 2500,  nCores = 8)
\end{verbatim}
The \textVerb{print} method of this object gives an overview of characteristics of the sample network (e.g., the number of estimated edges) and tips for further investigation, such as how to plot the estimated sample network or any of the bootstrapped networks. The \textVerb{summary}  method can be used to create a summary table of certain statistics containing quantiles of the bootstraps. 

\begin{figure*}
\centering
    \includegraphics[width=1\textwidth]{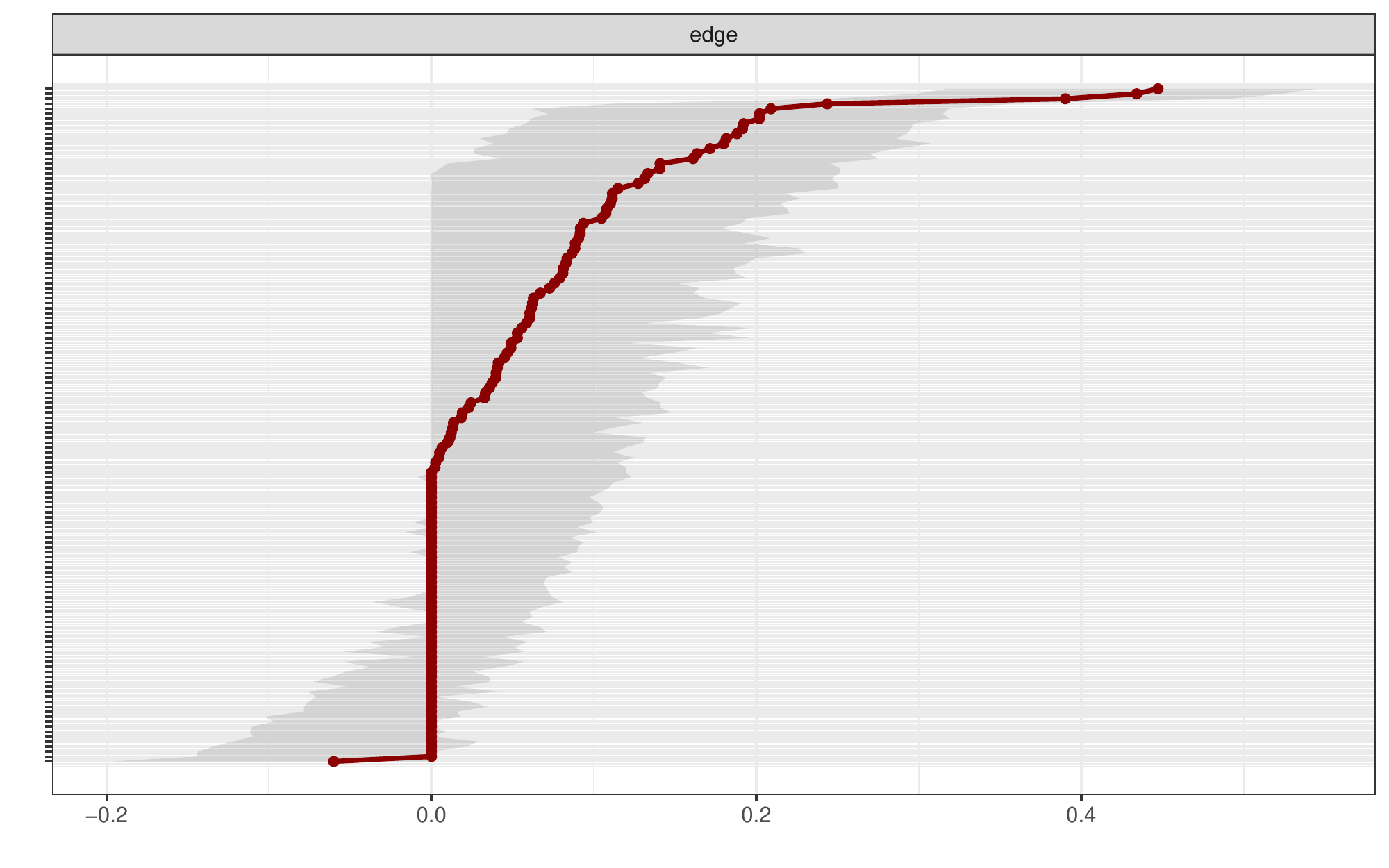}
\caption{Bootstrapped confidence intervals of estimated edge-weights for the estimated network of 17 PTSD symptoms. The red line indicates the sample values and the gray area the bootstrapped CIs. Each horizontal line represents one edge of the network, ordered from the edge with the highest edge-weight to the edge with the lowest edge-weight. In the case of ties (for instance, multiple edge-weights were estimated to be exactly 0), the mean of the bootstrap samples was used in ordering the edges. y-axis labels have been removed to avoid cluttering.}
\label{bootnet:fig:edgeCI}
\end{figure*}

The \textVerb{plot} method can be used to show the bootstrapped CIs for estimated edge parameters:
\begin{verbatim}
plot(boot1, labels = FALSE, order = "sample")
\end{verbatim}
Figure~\ref{bootnet:fig:edgeCI} shows the resulting plots and reveals sizable bootstrapped CIs around the estimated edge-weights, indicating that many edge-weights likely do not significantly differ from one-another. The generally large bootstrapped CIs imply that interpreting the order of most edges in the network should be done with care. Of note, the edges 16 (upset when reminded of the trauma) -- 17 (upsetting thoughts/images), 3 (being jumpy) -- 4 (being alert) and 5 (feeling distant) -- 11 (loss of interest), are reliably the three strongest edges since their bootstrapped CIs do not overlap with the bootstrapped CIs of any other edges.\footnote{As with any CI, non-overlapping CIs indicate two statistics significantly differ at the given significance level. The reverse is not true; statistics with overlapping CIs might still significantly differ.}

\begin{figure*}
\centering
    \includegraphics[width=1\textwidth]{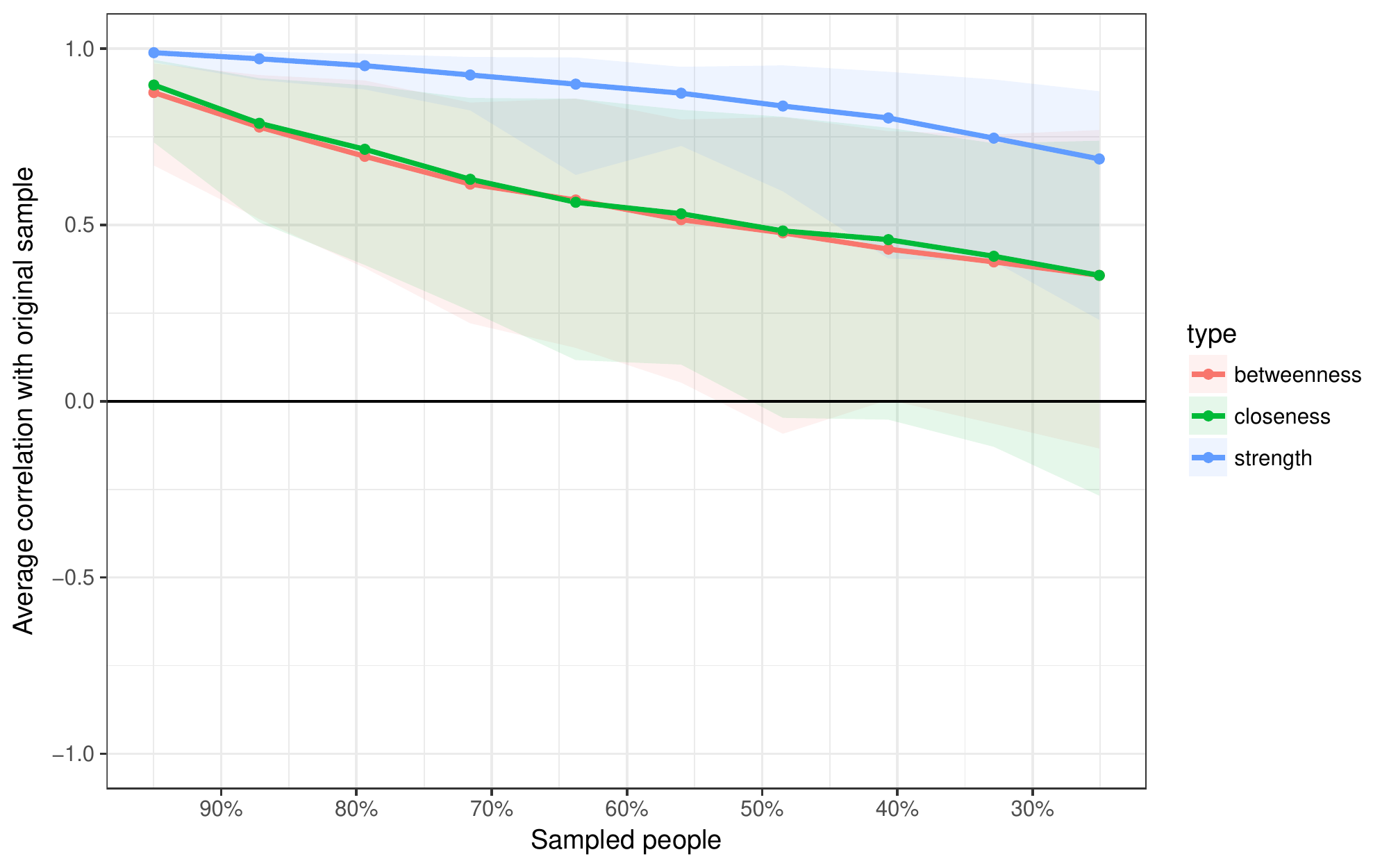}
\caption{Average correlations between centrality indices of networks sampled with persons dropped and the original sample. Lines indicate the means and areas indicate the range from the $2.5$th quantile to the $97.5$th quantile.}
\label{bootnet:fig:caseDrop}
\end{figure*}

\paragraph{Centrality stability} We can now investigate the stability of centrality indices by estimating network models based on subsets of the data. The case-dropping bootstrap can be used by using \textVerb{type = "case"}:
\begin{verbatim}
boot2 <- bootnet(Network, nBoots = 2500, type = "case", nCores = 8)
\end{verbatim}
To plot the stability of centrality under subsetting, the plot method can again be used:
\begin{verbatim}
plot(boot2)
\end{verbatim}
Figure~\ref{bootnet:fig:caseDrop} shows the resulting plot: the stability of closeness and betweenness drop steeply while the stability of node strength is better. This stability can be quantified using the $CS$-coefficient, which quantifies the maximum proportion of cases that can be dropped to retain, with $95\%$ certainty, a correlation with the original centrality of higher than (by default) $0.7$. This coefficient can be computed using the \textVerb{corStability} function:
\begin{verbatim}
corStability(boot2)
\end{verbatim}
The $CS$-coefficient indicates that betweenness ($CS(\mathrm{cor} = 0.7) = 0.05$) and ($CS(\mathrm{cor} = 0.7) = 0.05$) closeness are not stable under subsetting cases. Node strength performs better ($CS(\mathrm{cor} = 0.7) = 0.44$), but does not reach the cutoff of $0.5$ from our simulation study required consider the metric stable. Therefore, we conclude that the order of node strength is interpretable with some care, while the orders of betweenness and closeness are not. 

\paragraph{Testing for significant differences} The \textVerb{differenceTest} function can be used to compare edge-weights and centralities using the bootstrapped difference test. This makes use of the non-parametric bootstrap results (here named \textVerb{boot1}) rather than the case-dropping bootstrap results. For example, the following code tests if Node~3 and Node~17 differ in node strength centrality:
\begin{verbatim}
differenceTest(boot1, 3, 17, "strength") 
\end{verbatim}
The results show that these nodes do not differ in node strength since the bootstrapped CI includes zero (CI: $-0.20, 0.35$). The plot method can be used to plot the difference tests between all pairs of edges and centrality indices. For example, the following code plots the difference tests of node strength between all pairs of edge-weights:
\begin{verbatim}
plot(boot1, "edge", plot = "difference", onlyNonZero = TRUE,
     order = "sample")
\end{verbatim}
In which the \textVerb{plot} argument has to be used because the function normally defaults to plotting bootstrapped CIs for edge-weights, the \textVerb{onlyNonZero} argument sets so that only edges are shown that are nonzero in the estimated network, and \textVerb{order = "sample"} orders the edge-weights from the most positive to the most negative edge-weight in the sample network.  We can use a similar code for comparing node strength:
\begin{verbatim}
plot(boot1, "strength")
\end{verbatim}
In which we did not have to specify the \textVerb{plot} argument as it is set to the \textVerb{"difference"} by default when the statistic is a centrality index.

\begin{figure*}
\centering
\begin{subfigure}[b]{1\textwidth}
	  	\includegraphics[width=1\textwidth]{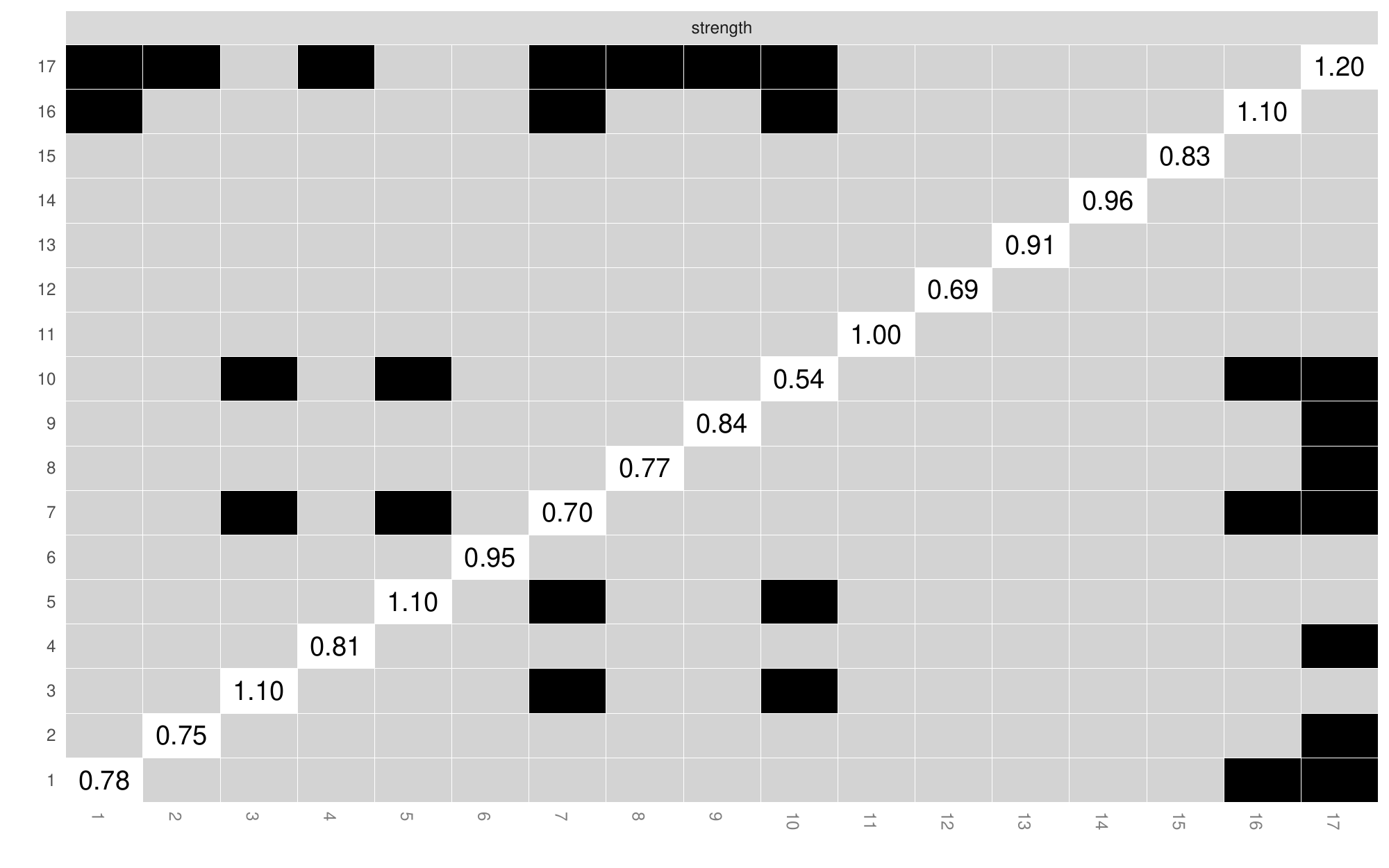}
        %\caption{Significant differences in node strength}
    \end{subfigure} \\
    \vspace{1cm}
 \begin{subfigure}[b]{1\textwidth}
	  	\includegraphics[width=1\textwidth]{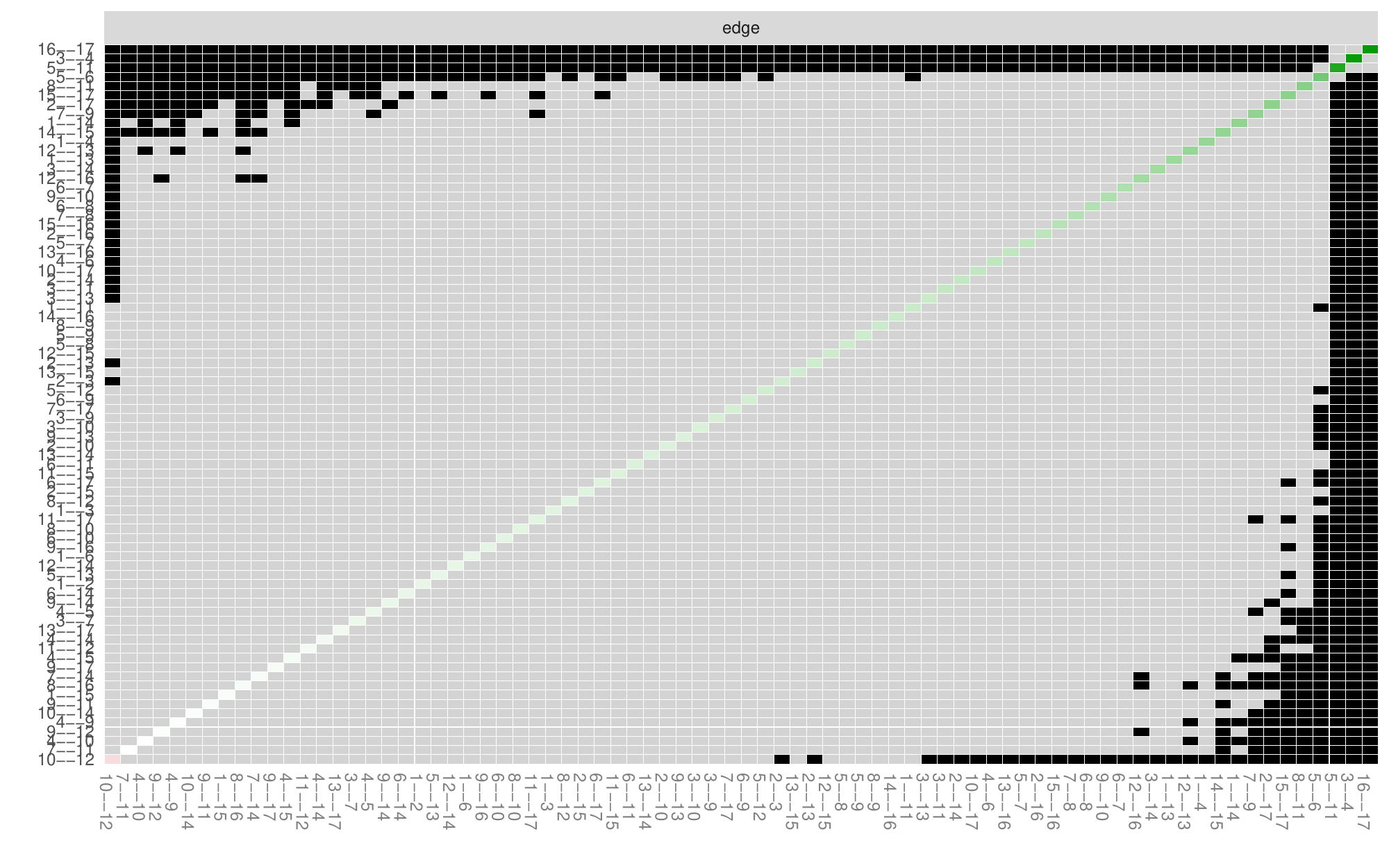}
       % \caption{Significant differences in edge-weights}
    \end{subfigure} 
\caption{Bootstrapped difference tests ($\alpha = 0.05$) between edge-weights that were non-zero in the estimated network (above) and node strength of the 17 PTSD symptoms (below). Gray boxes indicate nodes or edges that do not differ significantly from one-another and black boxes represent nodes or edges that do differ significantly from one-another.  Colored boxes in the edge-weight plot correspond to the color of the edge in Figure~\ref{bootnet:fig:net}, and white boxes in the centrality plot show the value of node strength.}
\label{bootnet:fig:sigdif}
\end{figure*}

The resulting plots are presented in Figure~\ref{bootnet:fig:sigdif}. The top panel shows that many edges cannot be shown to significantly differ from one-another, except for the previously mentioned edges 16 (upset when reminded of the trauma) -- 17 (upsetting thoughts/images), 3 (being jumpy) -- 4 (being alert) and 5 (feeling distant) -- 11 (loss of interest), which significantly differ from most other edges in the network. The bottom panel shows that most node strengths cannot be shown to significantly differ from each other. The node with the largest strength, Node~17, is significantly larger than almost half the other nodes. Furthermore, Node~7 and Node~10 and also feature node strength that is significantly larger than some of the other nodes. In this dataset, no significant differences were found between nodes in both betweenness and closeness (not shown). For both plots it is important to note that \emph{no} correction for multiple testing was applied.

\section{Simulation Studies}

We conducted three simulation studies to assess the performance of the methods described above. In particular, we investigated the performance of (1) the $CS$-coefficient and the bootstrapped difference test for (2) edge-weights and (3) centrality indices. All simulation studies use networks of 10 nodes. The networks were used as partial correlation matrices to generate multivariate normal data, which were subsequently made ordinal with four levels by drawing random thresholds; we did so because most prior network papers estimated networks on ordinal data (e.g., psychopathological symptom data). We varied sample size between $100$, $250$, $500$, $1{,}000$, $2{,}500$ and $5{,}000$, and replicated every condition 1,000 times. We estimated Gaussian graphical models, using the graphical LASSO in combination with EBIC model selection \citep{primerpaper,foygel2010extended}, using polychoric correlation matrices as input. Each bootstrap method used $1{,}000$ bootstrap samples. In addition, we replicated every simulation study with 5-node and 20-node networks as well, which showed similar results and were thus not included in this paper to improve clarity.

%%% UNTILL HERE %%%

\paragraph{$CS$-coefficients} We assessed the $CS$-coefficient in a simulation study for two cases where: networks where centrality did not differ between nodes, and networks where centrality did differ. We simulated chain networks as shown in Figure~\ref{bootnet:fig:1} consisting of 10 nodes, $50\%$ negative edges and all edge-weights set to either $0.25$ or $-0.25$. Next, we randomly rewired edges as described by \citet{watts1998} with probability $0$, $0.1$, $0.5$ or $1$. A rewiring probability of $0.5$ indicates that every edge had a $50\%$ chance of being rewired to another node, leading to a different network structure than the chain graph. This procedure creates a range of networks, ranging from chain graphs in which all centralities are equal (rewiring probability = 0) to random graphs in which all centralities may be different (rewiring probability = 1). Every condition (rewiring probability $\times$ sample size) was replicated $1{,}000$ times, leading to $24{,}000$ simulated datasets. On each of these datasets, case-dropping bootstrap was performed and the $CS$-coefficient was computed. Case-dropping bootstrap used $5{,}000$ bootstrap samples and tested $25$ different sampling levels (rather than the default $1{,}000$ bootstrap samples and $10$ different sampling levels) to estimate the $CS$-coefficient with more accuracy. Figure 7 shows the results, showing that the $CS$-coefficient remains low in networks in which centrality does not differ and rises as a function of sample size in networks in which centralities do differ. It can be seen that under a model in which centralities do not differ the $CS$-coefficient remains stable as sample size increases and stays mostly below .5, and roughly $75\%$ stays below $0.25$. Therefore, to interpret centrality differences the $CS$-coefficient should not be below $0.25$, and preferably above $0.5$.

\begin{figure*}
\centering
\includegraphics[width=1\textwidth]{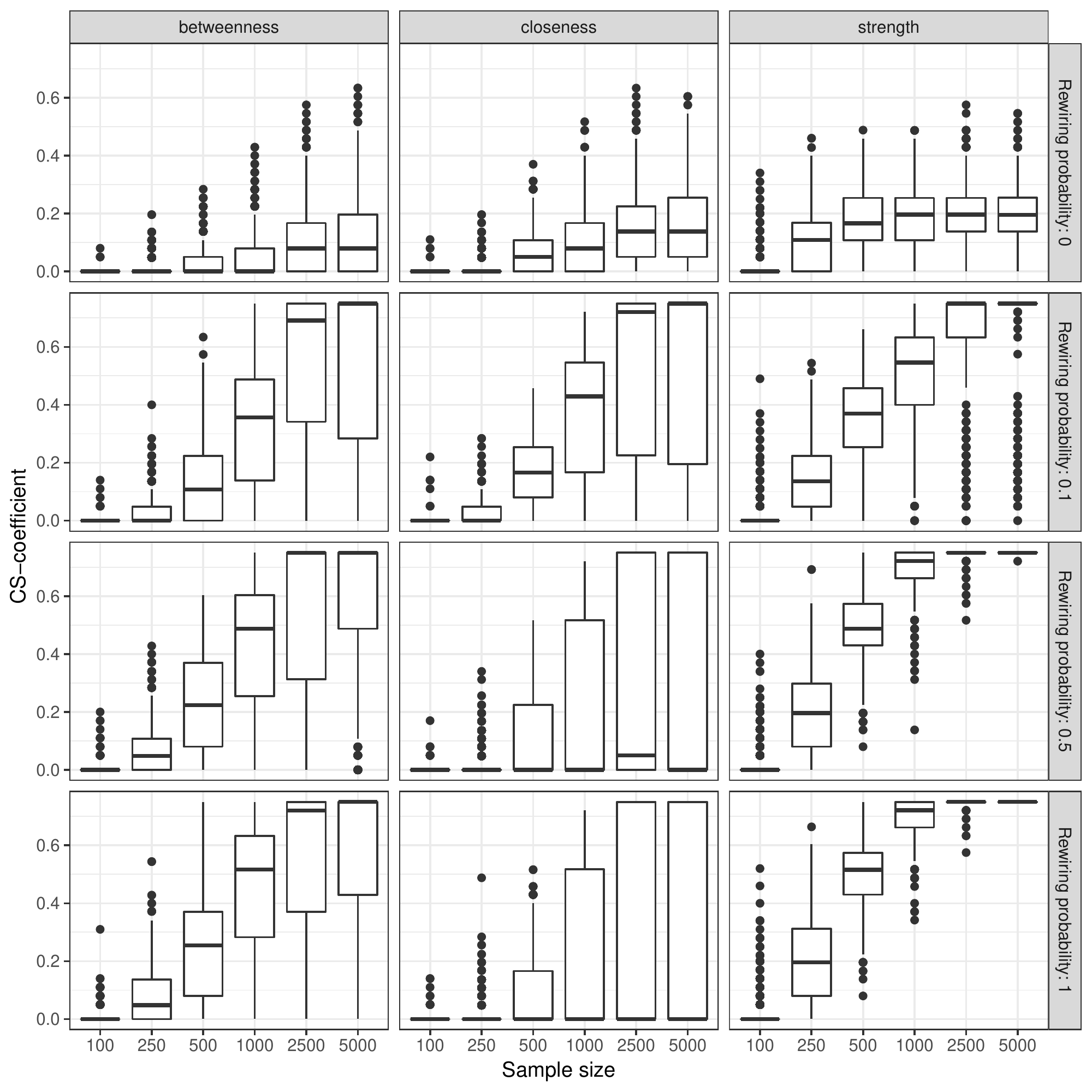}
\caption{Simulation results showing the $CS$-coefficient of $24{,}000$ simulated datasets. Datasets were generated using chain networks (partial correlations) of 10 nodes with edge-weights set to $0.25$ or $-0.25$. Edges were randomly rewired to obtain a range from networks ranging from networks in which all centralities are equal to networks in which all centralities differ. The $CS$-coefficient quantifies the maximum proportion of cases that can be dropped at random to retain, with $95\%$ certainty, a correlation of at least $0.7$ with the centralities of the original network. Boxplots show the distribution of $CS$-coefficients obtained in the simulations. For example, plots on top indicate that the $CS$-coefficient mostly stays below $0.2$ when centralities do not differ from one-another (chain graph as shown in Figure~\ref{bootnet:fig:1}).}
\label{bootnet:fig:CScoef}
\end{figure*}

\begin{figure*}
\centering
\includegraphics[width=1\textwidth]{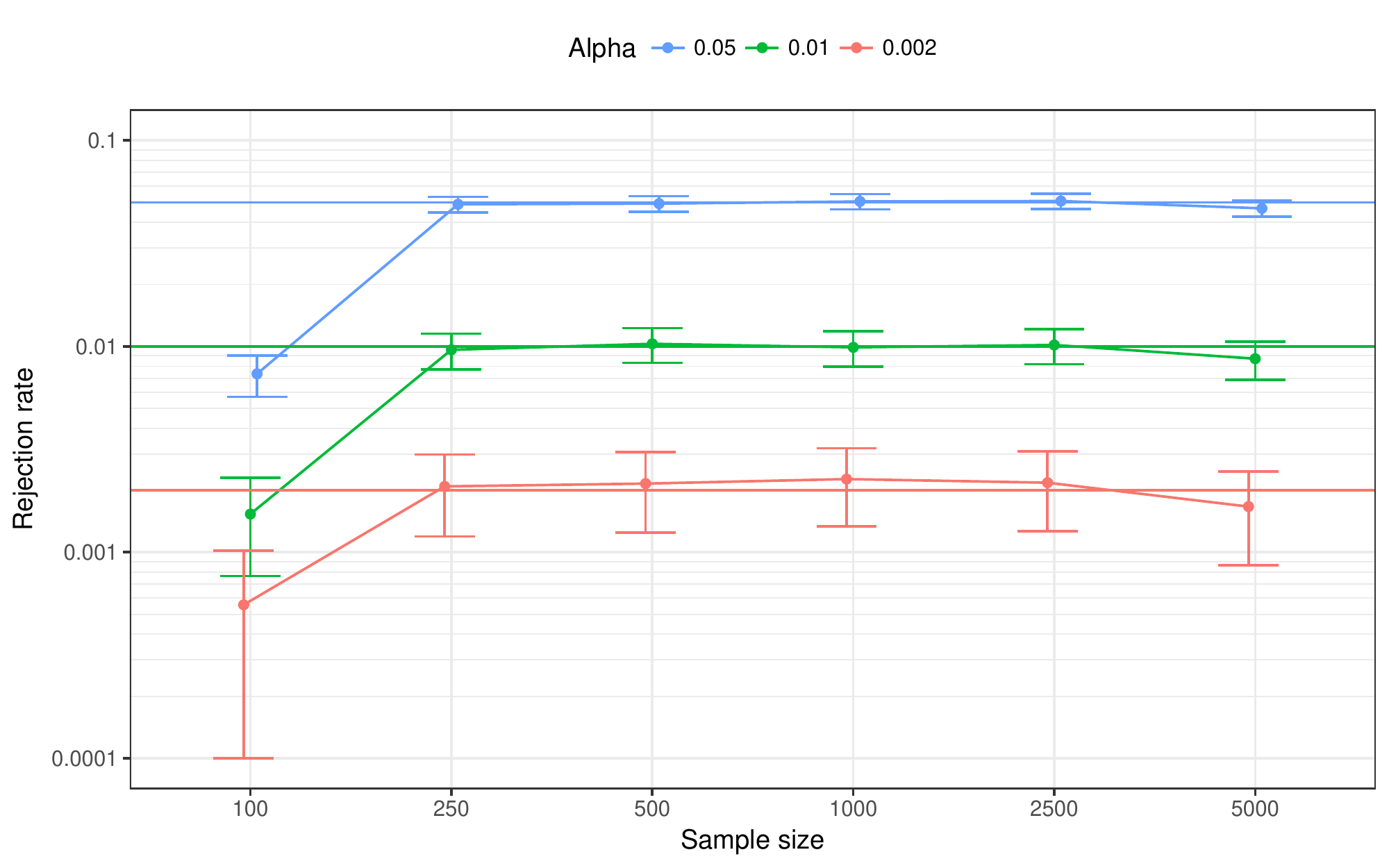}
\caption{Simulation results showing the rejection rate of the bootstrapped difference test for edge-weights on $6{,}000$ simulated datasets. Datasets were generated using chain networks (partial correlations) of 10 nodes with edge-weights set to $0.3$. Only networks that were nonzero in the true network were compared to one-another. Lines indicate the proportion of times that two random edge-weights were significantly different (i.e., the null-hypothesis was rejected) and their CI (plus and minus $1.96$ times the standard error). Solid horizontal lines indicate the intended significance level and horizontal dashed line the expected significance level given $1{,}000$ bootstrap samples. The $y$-axis is drawn using a logarithmic scale.}
\label{bootnet:fig:rejectRate}
\end{figure*}

\paragraph{Edge-weight bootstrapped difference test} We ran a second simulation study to assess the performance of the bootstrapped difference test for edge-weights. In this simulation study, chain networks were constructed consisting of 10 nodes in which all edge-weights were set to $0.3$. Sample size was again varied between $100$, $250$, $500$, $1{,}000$, $2{,}500$ and $5{,}000$ and each condition was again replicated $1{,}000$ times, leading to $6{,}000$ total simulated datasets. Data were made ordinal and regularized partial correlation networks were estimated in the same manner as in the previous simulation studies. We used the default of $1{,}000$ bootstrap samples to compare edges that were nonzero in the true network (thus, edges with a weight of $0.3$ that were not different from one-another), and investigated the rejection rate under different levels of $\alpha$: $0.05$, $0.01$ and $0.002$ (the minimum $\alpha$ level when using $1{,}000$ bootstrap samples). Figure~\ref{bootnet:fig:rejectRate} shows that rejection rate converged on the expected rejection rate with higher samples, and was lower than the expected rejection rate in the low sample condition of $N = 100$---a result of the LASSO pulling many edge-weights to zero in low sample sizes. 

\begin{figure*}
\centering
\includegraphics[width=1\textwidth]{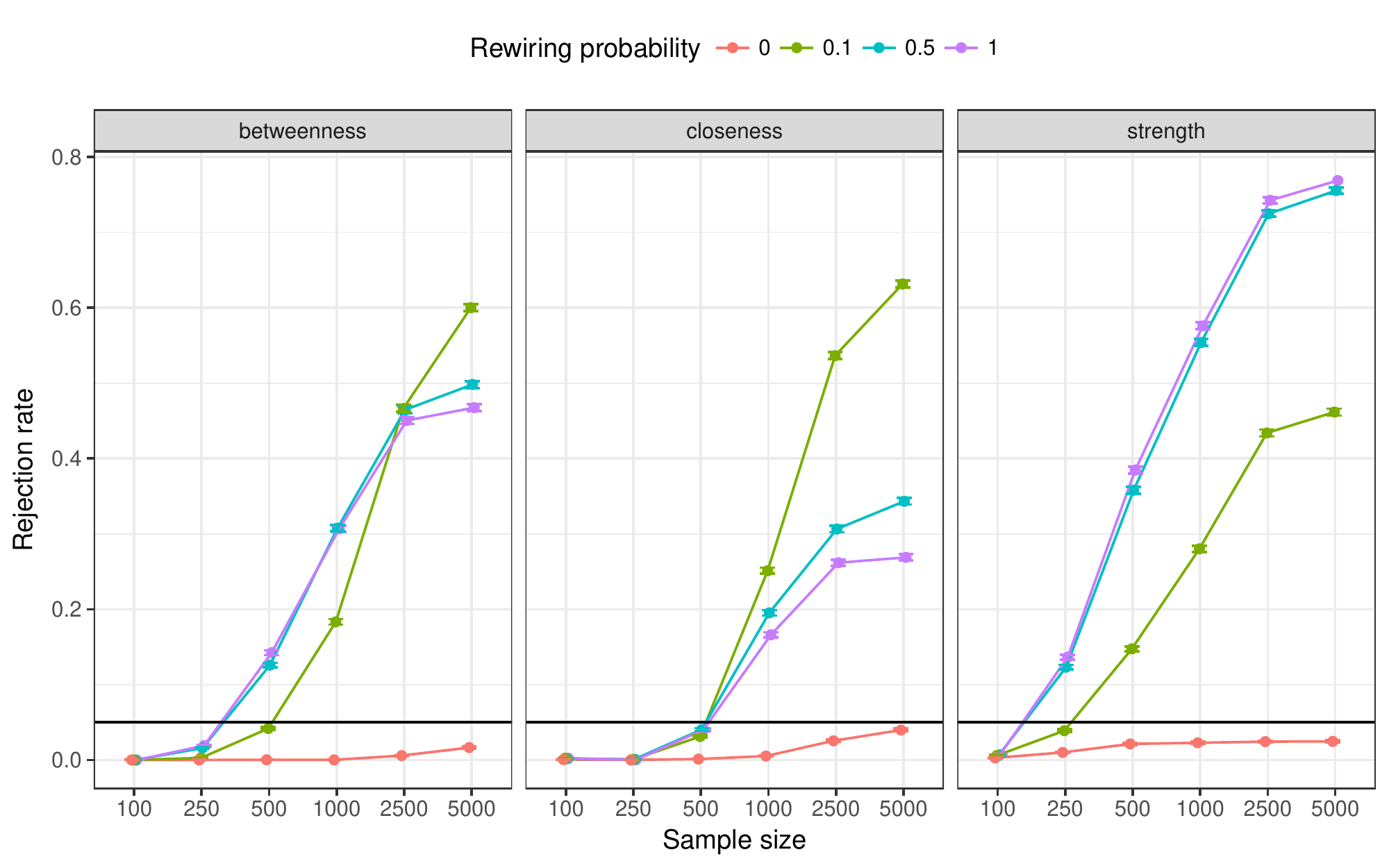}
\caption{Simulation results showing the rejection rate of the bootstrapped difference test for centrality indices. Datasets were generated using the same design as in Figure~\ref{bootnet:fig:CScoef}. Lines indicate the proportion of times that two random centralities were significantly different (i.e., the null-hypothesis was rejected at $\alpha = 0.05$).}
\label{bootnet:fig:simCent}
\end{figure*}

\paragraph{Centrality bootstrapped difference test} We conducted a third simulation study to assess the performance of the bootstrapped difference test for centrality indices. The design was the same as the first simulation study, leading to $24{,}000$ total simulated datasets. We performed the bootstrapped difference test, using $1{,}000$ bootstrap samples and $\alpha = 0.05$, to all pairs of nodes in all networks and computed the rate of rejecting the null-hypothesis of centralities being equal. Figure~\ref{bootnet:fig:simCent} shows the results of this simulation study. It can be seen that the average rate of rejecting the null-hypothesis of two centrality indices being equal under a chain-network such as shown in Figure~\ref{bootnet:fig:1} stays below 0.05 at all sample sizes for all centrality indices. As such, checking if zero is in the bootstrapped CI on differences between centralities is a valid null-hypothesis test. Figure~\ref{bootnet:fig:simCent}, however, also shows that the rejection rate often is below $0.05$, leading to a reduced power in the test. As such, finding true differences in centrality might require a larger sample size. When centralities differ (rewiring probability $> 0$), power to detect differences goes up as a function of sample size. Unreported simulation studies showed that using Pearson or Spearman correlations on ordinal data using this method leads to an inflated Type-I error rate. Our simulations thus imply that bootstrapped difference test for centrality indices for ordinal data should use polychoric correlations as input to the graphical LASSO.

\section{Discussion}

In this paper, we have summarized the state-of-the-art in psychometric network modeling, provided a rationale for investigating how susceptible estimated psychological networks are to sampling variation, and described several methods that can be applied after estimating a network structure to check the accuracy and stability of the results. We proposed to perform these checks in three steps: (A) assess the \emph{accuracy} of estimated edge-weights,  (B) assess the \emph{stability} of centrality indices after subsetting the data, and (C) test if edge-weights and centralities differ from one-another. Bootstrapping procedures can be used to perform these steps. While bootstrapping edge-weights is straight-forward, we also introduced two new statistical methods: the \emph{correlation stability coefficient} ($CS$-coefficient) and the \emph{bootstrapped difference test} for edge-weights and centrality indices to aid in steps 2 and 3 respectively. To help researchers conduct these analyses, we have developed the freely available R package \emph{bootnet}, which acts as a generalized framework for estimating network models as well as performs the accuracy tests outlined in this paper. It is of note that, while we demonstrate the functionality of \emph{bootnet} in this tutorial using a Gaussian graphical model, the package can be used for any estimation technique in R that estimates an undirected network (such as the Ising model with binary variables).

\paragraph{Empirical example results} The accuracy analysis of a 17-node symptom network of 359 women with (subthreshold) PTSD showed a network that was susceptible to sampling variation. First, the bootstrapped confidence intervals of the majority of edge-weights were large. Second, we assessed the stability of centrality indices under dropping people from the dataset, which showed that only node strength centrality was moderately stable; betweenness and closeness centrality were not. This means that the order of node strength centrality was somewhat interpretable, although such interpretation should be done with care. Finally, bootstrapped difference tests at a significance level of $0.05$ indicated that only in investigating node strength could statistical differences be detected between centralities of nodes, and only three edge-weights were shown to be significantly higher than most other edges in the network.

\subsection{Limitations and Future Directions}

\paragraph{Power-analysis in psychological networks} Overall, we see that networks with increasing sample size are estimated more accurately. This makes it easier to detect differences between centrality estimates, and also increases the stability of the order of centrality estimates. But how many observations are needed to estimate a reasonably stable network? This important question usually referred to as \emph{power-analysis} in other fields of statistics \citep{cohen1977statistical} is largely unanswered for psychological networks. When a reasonable prior guess of the network structure is available, a researcher might opt to use the \emph{parametric} bootstrap, which has also been implemented in bootnet, to investigate the expected accuracy of edge-weights and centrality indices under different sample sizes. However, as the field of psychological networks is still young, such guesses are currently hard to come by. As more network research will be done in psychology, more knowledge will become available on graph structure and edge-weights that can be expected in various fields of psychology. As such, power calculations are a topic for future research and are beyond the scope of the current paper.

\paragraph{Future directions} While working on this project, two new research questions emerged: is it possible to form an unbiased estimator for centrality indices in partial correlation networks, and consequently, how should true $95\%$ confidence intervals around centrality indices be constructed? As our example highlighted, centrality indices can be highly unstable due to sampling variation, and the estimated sampling distribution of centrality indices can be severely biased. At present, we have no definite answer to these pressing questions that we discuss in some more detail in the Supplementary Materials. In addition, constructing bootstrapped CIs on very low significance levels is not feasible with a limited number of bootstrap samples, and approximating $p$-values on especially networks estimated using regularization is problematic. As a result, performing difference tests while controlling for multiple testing is still a topic of future research. Given the current emergence of network modeling in psychology, remediating these questions should have high priority.

\paragraph{Related research questions} We only focused on accuracy analysis of cross-sectional network models. Assessing variability on longitudinal and multi-level models is more complicated and beyond the scope of current paper; it is also not implemented in \emph{bootnet} as of yet. We refer the reader to Bringmann and colleagues (\citeyear{bringmann2015revealing}) for a demonstration on how confidence intervals can be obtained in a longitudinal multi-level setting. We also want to point out that the results obtained here may be idiosyncratic to the particular data used. In addition, it is important to note that the bootstrapped edge-weights should not be used as a method for comparing networks based on different groups, (e.g., comparing the bootstrapped CI of an edge in one network to the bootstrapped CI of the same edge in another network) for which a statistical test is being developed.\footnote{\url{http://www.github.com/cvborkulo/NetworkComparisonTest}}  Finally, we wish to point out promising research on obtaining exact $p$-values and confidence intervals based on the results of LASSO regularized analyses (see \citealt{hastie2015statistical}, for an overview), which may in the future lead to a lesser need to rely on bootstrapping methods. 

\subsection{Conclusion}

In addition to providing a framework for network estimation as well as performing the accuracy tests proposed in this paper, \emph{bootnet} offers more functionality to further check the accuracy and stability of results that were beyond the scope of this paper, such as the parametric bootstrap, node-dropping bootstrap \citep{costenbader2003stability} and plots of centrality indices of each node under different levels of subsetting. Future development of \emph{bootnet} will be aimed to implement functionality for a broader range of network models, and we encourage readers to submit any such ideas or feedback to the Github Repository.\footnote{\url{http://www.github.com/sachaepskamp/bootnet}} Network accuracy has been a blind spot in psychological network analysis, and the authors are aware of only one prior paper that has examined network accuracy \citep{fried2016}, which used an earlier version of \emph{bootnet} than the version described here. Further remediating the blind spot of network accuracy is of utmost importance if network analysis is to be added as a full-fledged methodology to the toolbox of the psychological researcher.

% Non-BibTeX users please use
\bibliographystyle{apalike}
\bibliography{Bibliography}
%\bibliographystyle{bpacite}
%\bibliography{Bibliography}

\end{document}